\documentclass[twoside,10pt,a4paper]{newFNLstyle}
\usepackage{graphics}
\usepackage{cite}

\begin{document}

\volnumpagesyear{2}{4}{L279-L284}{2002}
\dates{30 May 2002}{15 October 2002}{6 November 2002}

\title{\vspace{-1.2truecm} SURVIVAL OF THE FITTEST AND ZERO SUM GAMES%
\thanks{Special Issue on Game Theory and Evolutionary Processes:
Order from Disorder---The Role of Noise in Creative Processes,
eds. D. Abbott, P.C.W. Davies and C.R. Shalizi, {\tt quant-ph/0206014}.}
}

\authorsone{APOORVA PATEL}
\affiliationone{CTS and SERC, Indian Institute of Science,
                Bangalore-560012, India}
\mailingone{adpatel@cts.iisc.ernet.in}

\maketitle

\markboth{A. Patel}{Survival of the Fittest and Zero Sum Games}

\pagestyle{myheadings}
% Comment this out to remove the running heads

\keywords{Darwinian evolution; Markov chain; quantum algorithms.}
% Keywords have to before the abstract I'm afraid.

\begin{abstract}
Competition for available resources is natural amongst coexisting species,
and the fittest contenders dominate over the rest in evolution. The dynamics
of this selection is studied using a simple linear model. It has similarities
to features of quantum computation, in particular conservation laws leading
to destructive interference. Compared to an altruistic scenario, competition
introduces instability and eliminates the weaker species in a finite time.
\end{abstract}

\section{Darwinian Evolution}

Charles Darwin explained the evolution of living organisms on the basis of
the observation that only the species best adapted to their surroundings
survive in a competitive environment. This idea, paraphrased as ``survival
of the fittest'', has since become the cornerstone of evolutionary theories.
The idea has been refined over the years, and we now understand its genetic
underpinnings. Hereditary transmission of genetic information is not perfect,
and occasional errors in DNA replication produce mutations of living organisms.
The mutated organism is in essentially the same environment as the original
one, and both have to compete for the available resources. The resources are
limited, and so the organism that can use them more efficiently obtains an
edge over its competitors. The net result is that if the mutation improves
the ability of the organism to survive, the mutated organism grows in number,
otherwise it fades away. In this mechanism, the mutations are not conscious
adaptations. They occur randomly---as a matter of fact most of them fail---but
once in a while they modify the organism in the right manner and improve the
chances of its survival. If the intermediate steps are glossed over, the
outcome looks like a solution to the optimisation problem, with the organisms
adapting to the selection pressures exerted by the environment. All this is
fairly logical, and can be illustrated by many examples (see for instance
\cite{dawkins}).
My aim here is to quantify this mechanism using a simple evolutionary model.

More specifically, I consider the situation where a number of species
are dependent on a common physical resource, say food. (This leaves out 
predator-prey conflicts where one species may become another's resource.)
When the resource is available in plenty, there is hardly any competition,
and all species prosper. The rate of growth of a species is then proportional
to its population, and so all populations grow exponentially. This typically
happens in the early stage of evolution of a new species. No resource is
unlimited, however, even if it is continuously regenerated (e.g. food
availability ultimately depends on sunlight). A time comes when competition
for the resource is inevitable. In the later stage of evolution, therefore,
one species can prosper only at the expense of another, and eventually
populations stabilise.

Such a situation can be described by a set of linear evolution equations.
(Quad-ratic evolution equations are more appropriate in a predator-prey
setting.) Let the index $i$ label a set of coexisting species in a given
environment, and let $\phi_i(t)$ denote their populations at time $t$.
Then the discrete time evolution of the populations can be written as a
matrix equation (generically $M$ is not symmetric),
\begin{equation}
\phi_i(t+\Delta t) = \sum_j M_{ij}(\Delta t)~\phi_j(t) ~,~~ \phi_i(t) \ge 0 ~.
\end{equation}
The fact that the next generation populations must arise from the present
generation ones, even in case of mutations, makes this equation homogeneous.
The diagonal terms $M_{ii}$ represent the individual rates of growth, while
the off-diagonal terms $M_{i\ne j}$ represent interactions between species.
As mentioned above, when the resource is available in plenty, the populations
evolve according to:
\begin{equation}
M_{i\ne j} \approx 0 ~\Longrightarrow~
\phi_i(t) \approx (M_{ii})^{t/\Delta t} \phi_i(t=0) ~.
\end{equation}
The interesting situation is the competitive stage, where the populations
become interdependent due to mutual interactions and the nature of the
interactions (para-metrised by $M_{i\ne j}$) determines how they evolve.

\section{Zero Sum Games}

To analyse the competitive stage, let us choose normalisations such that
a unit population of any species consumes the same amount of resource,
and the total resource available at any time is one. Then we have the
conservation laws:
\begin{equation}
\sum_i \phi_i = 1 ~,~~ \sum_i M_{ij} = 1 ~.
\end{equation}
The constraint on $\phi_i$ arises because the total population supported by
a fixed resource is fixed. When applied to both sides of Eq.(1), it yields
the constraint on $M_{ij}$ as a consistency condition. The continuity of
evolution in time implies that
\begin{equation}
M_{ij} = \delta_{ij} + C_{ij} ~,~~ C_{ij} = {\cal O}(\Delta t) ~,~~
\sum_i C_{ij} = 0 ~.
\end{equation}
The matrix $C_{ij}$, describing the change in populations, clearly shows
that what is gained by some species is lost by some others. Obviously
$M_{i\ne j}$ can take positive as well as negative values. As a matter of
fact, situations of both positive and negative interactions occur routinely
in biological systems (e.g. catalysis and inhibition, symbiosis and parasitic
behaviour, defence mechanisms and cancer, etc.).

Situations of this type, where there is no net gain or loss, have been
formally called zero sum games. In the conventional form of zero sum games,
a multitude of strategies are available to the players, and competition
amongst the players leads them to choose a stable strategy. This stable
strategy turns out to be a ``mini-max solution'', i.e. each player chooses
the strategy that maximises his minimum gain \cite{vonneumann}.
No player can improve his performance by unilaterally departing from this
stable strategy, and the explicit solution can be obtained by linear
programming methods.

The model described by Eqs.(1) and (3) is somewhat different. First, in the
evolutionary context, the strategies are not actively chosen but are created
by random mutations. Also competition for the resource is guaranteed,
because mutations produce closely related species in the same environment.
It is then of importance to study the dynamics of the system to understand
how the incorrect strategies are eliminated and the correct ones are selected.
That requires not just the evaluation of the stable solution but also the
manner in which it is approached.

Second, the quantitative gains and losses of various species depend not only
on the choice of strategy but also on their current populations. In a sense,
the more populous species gets more chances to grab the resources, and that
alters the final outcome of the competition.

\section{Stochastic Evolution}

Let us first look at the simpler situation when $0 \le M_{ij} \le 1$.
Eqs.(1) and (3) then describe stochastic evolution of probability
distributions, which has been extensively studied (see for instance
\cite{feller}).
The linear evolution is a Markov chain, which in the generic case is ergodic
and converges to a unique stationary distribution. (If the matrix $M$ has a
block-diagonal structure, then each block can be studied independently.) To
see this, consider the asymptotic behaviour of $M^n$ as $n\rightarrow\infty$.
At every iteration, $(M^{l+1})_{ij}$ is the weighted average of $(M^l)_{ik}$
$(k=1,\ldots,N)$ with weights $M_{kj}$. The process of repeated averaging
converges to the fixed point,
\begin{equation}
\lim_{n\rightarrow\infty} (M^n)_{ij} = m_i ~,~~ \sum_i m_i = 1 ~.
\end{equation}
With all rows identical, $\lim_{n\rightarrow\infty} M^n$ has only one
non-vanishing eigenvalue, equal to one. It follows that $M$ has the leading
eigenvalue $\lambda_1 = 1$ corresponding to the stationary distribution,
and remaining eigenvalues $|\lambda_{p\ne1}| < 1$ corresponding to transient
distributions. The left and right eigenvectors for the leading eigenvalue are,
\begin{equation}
(e^L_1)_i = 1/N ~,~~ (e^R_1)_i = m_i ~.
\end{equation}
The left and right eigenvectors satisfy the orthogonality relation,
$\sum_i (e^L_p)_i (e^R_q)_i = 0$ for $p \ne q$. All transient distributions
are orthogonal to $e^L_1$ and necessarily contain negative components.
The next-to-leading eigenvalue $\lambda_2$ provides the rate of convergence
towards the stationary distribution, and can be obtained my maximising
$\sum_{ij} x_i M_{ij} y_j$, with the vectors $x$ and $y$ orthogonal to
$e^R_1$ and $e^L_1$ respectively, and $\sum_i x_i y_i =1$. The convergence
is monotonic because $M_{ij} \ge 0$, $\phi_i \ge 0$.

All these properties belong to altruistic evolution, since $C_{i\ne j}\ge0$
while $C_{ii}\le0$. In the competitive case, some of the $M_{i\ne j}<0$,
and the consequences need to be analysed. The presence of both positive and
negative contributions produces cancellations. Monotonic convergence is no
longer automatic; oscillations and instabilities may occur in stead. An
important restriction is imposed by the physical requirement that no population
can become negative. (Note that there is no such requirement on the components
of the individual eigenvectors of $M$.) The evolution must be modified
whenever it drives some $\phi_i$ negative. The correct procedure is to stop
the evolution at the instance $\phi_i$ becomes zero, eliminate $i^{\rm th}$
row and column from the matrix $M$, and then continue evolution in the
reduced dimensional space. Obviously, $\phi_i$ can be driven negative only
if some $M_{i\ne j}$ is negative. The reduction of dimensionality, therefore,
decreases the number of negative $M_{ij}$ and increases the stability of the
system. The inverse process, increasing dimensionality of the space, occurs
when chance mutation creates a new species.

It is known that even with $0 \le M_{ij} \le 1$, Eq.(1) cannot be evolved
backward in time {\it indefinitely}. This is true in spite of the fact that
generic stochastic matrices are positive definite and $M^{-1}$ exists.
If backward evolution is attempted, some $\phi_i$ is driven negative at
some stage, and beyond that point interpretation of $\phi_i$ as populations
(or probabilities) is lost. Evolution further back in time is possible only
by modifying $M$. This is easy to see because eigenvalues of $M^{-1}$ are
the reciprocals of those for $M$, and with $|\lambda_{p\ne1}^{-1}| > 1$,
the growing transient distributions drive some $\phi_i$ out of their
physically allowed range $[0,1]$.

\section{Similarity to Quantum Computation}

It is useful to compare this evolution problem to quantum algorithms.
Both are constrained by conservation laws. While Eq.(3) preserves the
linear norm of the vector, quantum evolution preserves the quadratic norm.
The most general evolution preserving the quadratic norm is described by
orthogonal transformations for real variables, and by unitary transformations
for complex variables. Orthogonal transformations are generated by
antisymmetric matrices (generating matrices parametrise infinitesimal
group transformations in the neighbourhood of identity), and they inevitably
contain negative matrix elements---the negative elements are an automatic
consequence of the underlying conservation laws. (Only orthogonal
transformations without any negative matrix elements are simple permutation
matrices.) Although complete understanding of quantum evolution requires
use of complex numbers, classical language extended to include negative
probabilities can explain unusual features of certain simple systems (e.g.
quantum correlations of two spins violating Bell's inequalities) \cite{feynman}.
Recent developments in algorithms for quantum computation (see for instance
\cite{nielsen})
offer a hint of how evolution may change in presence of negative matrix
elements.

Classical algorithms based on Boolean logic can be expressed in terms of
permutation matrices. Quantum algorithms exploit two features to beat them,
superposition of states (which is a generic property of waves) and quantum
entanglement (which is not relevant here). Superposition means letting
multiple states be in the same place at the same time, and it can reduce the
spatial degrees of freedom required for the algorithm exponentially. Cleverly
designed destructive interference amongst superposed states can reduce the
the time required to execute the algorithm by eliminating unwanted states.
Coexistence of species in biological evolution is not quite the same as
superposition; it allows simultaneous evolution of all species but without
reducing the spatial degrees of freedom. But destructive interference is
still an option available to reduce the evolution time. The known quantum
algorithms reduce the execution time compared to their classical counterparts,
at least by a constant factor if not polynomially or exponentially. It is
important to note that biological evolution occurs over long time scales,
and even a tiny change in the rate of growth---fraction of a percent---matters
because that can translate into exponential changes in populations over a
long time. The lessons learnt from quantum computation thus suggest that
negative values of $M_{ij}$ may create destructive interference and help the
species reach their asymptotic populations faster, i.e. competition should
beat altruism in picking a winner amongst the contenders.

\section{Consequences of Destructive Interference}

Let us now explicitly analyse how evolution changes, when the range of
$M_{i\ne j}$ is extended to include negative values. Clearly, extending the
range of $M_{ij}$ cannot make the evolutionary process any less powerful.
The stationary eigenvalue of $M$, $\lambda_1=1$, which follows just from
the averaging properties as described above, remains unaffected. But other
eigenvalues and eigenvectors change, and modify evolution.
 
An example with two interacting species illustrates the possibilities. In
this case, the solution of Eq.(1) in terms of the (unnormalised) right
eigenvectors of the evolution matrix is:
\begin{equation}
M = \left( \matrix{1-\alpha &   \beta \cr
                     \alpha & 1-\beta \cr} \right) ~,~~
e^R_1 (\lambda_1 = 1) = \left( \matrix{\beta \cr \alpha \cr} \right) ~,~~
e^R_2 (\lambda_2 = 1-\alpha-\beta) = \left( \matrix{1 \cr -1 \cr} \right) ~,
\end{equation}
\begin{equation}
\phi(t=0) = \left( \matrix{a \cr 1-a \cr} \right) ~\Longrightarrow~
\phi(t) = \left( {1 \over \alpha+\beta} \right) e^R_1 + (\lambda_2)^{t/\Delta t}
          \left(a - {\beta \over \alpha+\beta} \right) e^R_2 ~.
\end{equation}
$\alpha,\beta$ are small, ${\cal O}(\Delta t)$. The evolutionary behaviour
depends on their signs :\\
(a) $\alpha\ge0,\beta\ge0$: The asymptotic population is proportional
to $e^R_1$. $\lambda_2<1$ and the transient part proportional to $e^R_2$
fades away. The two species coexist, and their population ratio is stable
against small perturbations. Such a behaviour does not occur in quantum
algorithms.\\
(b) $\alpha\beta<0$: Both $e^R_1$ and $e^R_2$ contain negative
components, and so the non-negative population vector has to be a mixture of
the two. In course of evolution, the population $\phi_i$ with $M_{i\ne j}<0$
is monotonically driven to zero. Thereafter evolution has to continue in the
reduced dimensional space.\\
(c) $\alpha\le0,\beta\le0$: The eigenvalue $\lambda_2>1$, and dominance
of $e^R_2$ drives one of the $\phi_i$ to zero. The relative size of initial
populations (i.e. comparison of $a/(1-a)$ vs. $\beta/\alpha$) determines
which $\phi_i$ is driven to zero. Afterwards evolution has to continue in
the reduced dimensional space. Such a dependence on the initial populations
is quite distinct from the ``mini-max analysis''.

These results for two interacting species have a topological interpretation.
The three regimes can be looked upon as minimisation of a function over an
interval, when it (a) has a single minimum, (b) is monotonic, and (c) has a
single maximum. The relation between linear evolution and gradient of a
quadratic form is generic, and the features exhibited in the above example
can be expected to generalise to more complicated multi-species systems.
Specifically, whenever some $M_{i\ne j}<0$ :\\
(1) The non-stationary eigenvectors play an important part in evolution.\\
(2) Having any $M_{i\ne j}<0$ makes the stationary eigenvector unstable, and
drives the system towards reduced dimensionality.\\
(3) Evolution one by one eliminates species $\phi_i$ with some $M_{ij}<0$,
till the reduced dimensional system no longer has any $M_{ij}<0$.\\
(4) The elimination of species takes place in finite time,
which depends on initial populations and $M_{ij}$, but is roughly
${\cal O}(\Delta t/C_{ij})$. This is in contrast to exponentially
decaying tails of transient parts in an altruistic evolution.\\
(5) The surviving population is given not by the stationary eigenvector of
the original system, but by the stationary eigenvector of the reduced
dimensional system.

The simple model presented above provides a quantification of features
anticipated in Darwinian evolution. The crucial ingredient has been the
limited availability of a resource leading to destructive interference.
Chance mutation may introduce a species that snatches away the resource
from another one. This always produces an instability, which eliminates
the weaker species in a finite time. The stronger survivors are stable
until the next mutation instability. It is worthwhile to observe that
the elementary components of biological systems are so simple and cheap
that they can be produced in large numbers even with limited resources.
As a result, biological systems often exhibit wastefulness, e.g. millions
of eggs and pollen grains are produced when a few would have sufficed to
propagate the species in a secure environment. Such an overkill actually
strengthens the competition and enforces survival of the fittest.

To summarise, the analysis presented in this article is straightforward,
and points out certain similarities between competitive evolution, zero-sum
games and quantum computation. The latter two possess a sound mathematical
framework, and so even a simple analogy with them can help us understand
better the behaviour of highly complex biological systems. The analogies
are not perfect, however, and the effect of the differences has to be
incorporated properly in the results. My conclusion is that limited
availability of resources leads to competition, which eliminates the weaker
species in a finite time. The arguments presented here can also be applied
to other competitive situations, e.g. economic and social interactions.

\end{document}